\documentclass{iopart}

\usepackage{iopams}

\usepackage{multicol}
\usepackage{graphicx}

\begin{document}

\title{Light Transport and Localization in Diffusive Random Lasers}

\author{
R. Frank, A. Lubatsch, J. Kroha
}

\address{
Physikalisches Institut and Bethe Center for Theoretical Physics,\\
Universit\"at Bonn, Nu{\ss}allee 12, 53115 Bonn, Germany
}
\begin{abstract}

We develop an analytical theory for diffusive random lasers by coupling the
transport theory of the disordered medium to the semiclassical laser rate 
equations, accounting for (coherent) stimulated and (incoherent) 
spontaneous emission. From the causality of wave propagation in an  
amplifying, diffusive medium we derive a novel length scale 
which we identify with the average mode radius of 
the lasing quasi-modes. We show further that loss at the surface of the 
laser-active medium is crucial in order to stabilize a stationary 
lasing state. The solution of the transport theory of random 
lasers for a layer geometry with appropriate surface boundary conditions 
yields the spatial profile of the light intensity and of the population
inversion. The dependence of the intensity correlation length on the 
pump rate is in qualitative agreement
with experimental and numerical findings.

\end{abstract}

\section{Introduction}

A random laser is a system formed by randomly distributed
scatterers embedded in a host medium where either scatterer or host
medium or both provide optical gain through stimulated 
emission \cite{Letokhov}. 
Recent observations of random lasing in a wide variety of systems, like
powders of semiconductor nanoparticles \cite{Markushev86,Cao98,Cao99,Kalt09}, 
organic dyes in strongly scattering media 
\cite{Lawandy94,Frolov99,Wiersma04}, organic films or 
nanofibers \cite{Anni04,Klein05,Quochi04} 
and ceramics \cite{Bahoura02},
have triggered a rapidly growing interest. For reviews with comprehensive  
lists of references see \cite{Cao03,Cao05}. 
Random lasers share some properties with conventional lasers, like
threshold behavior \cite{Cao99}, narrow spectral lines \cite{Cao00}, 
or photon statistics \cite{Cao01,Beenakker99}, but also exhibit 
distinctly different properties like multidirectional emission.
Coherent feedback has unambiguously been demonstrated to be present in
strongly disordered random lasers \cite{Cao01}. It requires the light
to be sufficiently confined in the random system. While spatially confined 
regions from which the laser emission takes place have been observed 
experimentally \cite{Cao00,Cao03}, the physical origin 
of coherent feedback, of localized quasimodes and the dependence of 
their size on the pump rate have remained controversial. The possible 
theoretical explanations
range from preformed random microresonators \cite{Shapiro02} to 
multiple random scattering (diffusion) \cite{Cao07}, possibly enhanced
by self-interference \cite{Cao00} of waves and the resulting onset of 
Anderson localization (AL) \cite{Anderson}. Conversely, it is an 
interesting fundamental question how AL of light, which has been 
understood\cite{gangof4,Woelfle80} as a consequence of self-interference,
is influenced by the coherent, stimulated amplification in the lasing
state. Problem of the intensity distribution in a diffusive random  
laser has been attacked theoretically by phenomenolocical 
diffusion models \cite{John04} and numerical calculations
in one  \cite{Soukoulis00,Cao04a} and two spatial
dimensions \cite{Cao01a,Cao04,Cao05a,Cao07}. However, the
experimentally observed decrease of the lasing spot size with 
increasing pump rate, has not been explained so far. 

In this paper, we address the question of the size of lasing spots 
in diffusive random lasers by an analytical transport theory. 
We begin the analysis in section 2 by an outline of the
transport theory  of light in a disordered medium with 
linear gain (i.e., constant-in-time amplification rate),
including self-interference (so-called Cooperon) contributions.
By a phenomenological analysis we show that causality implies in the 
presence of linear gain a novel length scale which is to
be identified by the average radius (spot size) of a lasing mode, $R_s$.
In section 3 we extend the theory for a system with linear gain to 
a transport theory including non-linear gain due to stationary 
lasing above threshold. 
Observing that a stationary lasing state is possible only if the
amplification in the medium is compensated by loss at the surface 
of the system, we consider a model for a finite-size
random laser with infinite extension in the ($x$,$y$) plane, but 
finite, constant thickness in the $z$ direction, a geometry relevant
for many experimental systems \cite{Cao98,Cao99,Cao00,Cao01}.
Coupling the diffusive transport theory to the rate equations of a 
four-level laser in the stationary state
we derive an analytical expression for the intensity correlation length
$\xi$ to be identified with the average 
lasing spot size, $R_s$. Due to the surface boundary conditions the
spatial extention of a lasing spot in the ($x$,$y$) plane 
obtains a z-dependent profile , $\xi(L)$. 
We also analyse its dependence on the pump rate.
The conclusions are drawn in section 4.
on the pump rate and on the depth of the mode along the $z$ direction.

\section{Transport Theory for a Diffusive, Linear-gain Medium and Causality}
\label{Sec_TT}

The propagation of light is described by its wave equation. 
Neglecting the polarization degree of freedom we consider 
in the following the scalar wave equation for the field $\Psi$.
It reads,
\begin{equation} 
\label{eq:field} 
\frac{\omega^2}{c^2} \, \epsilon (\vec{r}\,) \Psi_\omega(\vec{r}\,) 
+ \nabla ^2 \Psi_\omega( \vec{r}\,) 
= -i \omega \frac{4\pi}{c^2}  j_\omega(\vec{r}\,)\ , 
\end{equation} 
where $c$ denotes the vacuum speed of light  
and  $j_\omega (\vec{r}\,)$ an external current source. 
The dielectric constant is
$ \epsilon(\vec{r}\,) \!\!=\! \epsilon_b + \Delta\epsilon\, V(\vec{r}\,)$,  
where the dielectric contrast between the background, $\epsilon_b$, and the
scatterers, $\epsilon_s$, has been defined as
$\Delta\epsilon \!=\! \epsilon_s - \epsilon_b$. 
The spatial arrangement
of the scatterers is described through the function 
$V(\vec{r}\,)\!\! =\!\! \sum_{\vec{R}} S_{\vec{R}}\,(\vec{r}-\vec{R}\,)$, 
with  $S_{\vec{R}}\,(\vec{r}\,)$ a localized shape function 
at random locations $\vec{R}$. Linear gain (absorption) is described
by a temporally constant, negative (positive) imaginary part of 
$\epsilon_b$ and/or $\epsilon_s$.

In Refs. \cite{Lubatsch05,Frank06,Lubatsch09} we have developed 
a theory for light transport in disordered media with linear gain 
or absorption.
It results in an energy-density correlation function 
$P_{\mbox{\tiny E}}^{\omega}(\vec r-\vec r', t-t')$, which 
describes how the energy density of the light field with frequency
$\omega$ propagates diffusively between two points in space
and time, ($\vec r, t$), ($\vec r', t'$).
The Fourier transform of the energy-density correlation function 
$P_{\mbox{\tiny E}}^{\omega}(Q,\Omega)$ is obtained as
\begin{eqnarray} 
\label{P_E} 
P_{\mbox{\tiny E}}^{\omega}(Q,\Omega) = 
 \frac{N_{P}}
 {\Omega + i Q^2 D + i \xi_a^{-2}D}
\end{eqnarray}
where the expression for the coefficient $N_P$
is given explicitly in Ref. \cite{Frank06}, but is not relevant 
for the present purpose.
The denominator of Eq. (\ref{P_E}) exhibits the expected diffusion pole 
structure with the diffusion coefficient $D$. 
In addition, in the case of a non-conserving medium, i.e.
net absorption (gain), there appears the (purely imaginary) 
term $i\gamma_a= i\xi_a^{-2}D$, 
which has a positive (negative) imaginary part and does not
vanish in the hydrodynamic limit, $\Omega\to 0,\ Q \to 0$.  
The self-consistent solution of the transport theory including 
self-interference of waves (Cooperon contributions) 
(see Refs. \cite{Frank06,Lubatsch09}) 
shows that in the presence of absorption or gain the diffusion 
coefficient $D$ cannot vanish and is in general complex.
Hence, truly Anderson localized modes do not exist in this case.

For the case of absorption ($\gamma_a >0$) it is seen by
Fourier transforming Eq. (\ref{P_E}) w.r.t. time, 
$P_{\mbox{\tiny E}}^{\omega}(Q, t) = iN_P\, e^{-(Q^2D+\gamma_a)t}$, 
that $Re\gamma_a$ represents the loss rate of the photonic 
energy density due to absorption in the medium. 
Fourier transforming, on the other hand,  
Eq. (\ref{P_E}) w.r.t. space in the stationary limit 
($\Omega \to 0)$) shows that $\xi_a = Re \sqrt{(\gamma /D)}$ is the      
length scale over which the energy density of diffusive modes
is correlated in the lossy medium. 

For the case of linear gain ($\gamma_a <0$) the wave equation
predicts an unlimited growth of the field amplitude and, hence,
of the energy density. This means that a stationary lasing 
state is not possible in this case and, therefore, the limit
$\Omega \to 0$ must strictly not be taken in Eq. (\ref{P_E}).    
Such a behavior of linear gain is expected only during the
exponential intensity growth shortly after the onset of lasing. 
A complete theory of random lasing must, therefore, take into
account either the full temporal dynamics of the system, or 
in a stationary state additional surface loss effects must compesate 
for the gain in the medium (see section 3). 
Nevertheless, we can extract a characteristic size of a 
stationary lasing spot  
from this theory by requiring that the stationary lasing state has been
reached {\it locally}, i.e. within a finite subvolume of the
system: Causality requires that the pole of $P_E^{\omega}(Q,\Omega)$,
Eq. (\ref{P_E}),
as a function of $\Omega$ resides in the lower complex $\Omega$ 
half-plane. For $\gamma_a <0$ this is possible only if all the 
diffusive modes allowed inside a given lasing spot have a 
wavenumber $Q > Q_{min} = \sqrt{Re(-\gamma _a /D)}$. This, in turn,
requires that the spot size is 
\begin{eqnarray} 
\label{R_s} 
R_s = \frac{2\pi}{Q_{min}} = \frac{2\pi}{\sqrt{Re(-\gamma _a /D)}}\ . 
\end{eqnarray}
It is the characteristic, maximal size of a spatial region over which 
diffusive modes can be causally correlated in the stationary 
lasing state. We conjecture that, hence, this size is to be identified 
with the lasing spot size observed experimentally \cite{Cao00,Cao03} 
in diffusive random lasers. Since according to the microscopic theory 
\cite{Frank06} the growth rate $(-\gamma_a)$ is, for small linear
gain, proportional to the avergage gain in the medium, 
$\gamma_a \propto {\rm Im} \overline{\epsilon (\vec r)}$, 
we predict the spot size to be inversely proportional to the gain.  

More generally, despite the fact that the linear gain assumption is 
not suited to describe stationary lasing, it can be used to estimate 
the laser threshold, i.e. the critical pump rate for lasing. 
Amazingly, this is a rather general remark. For example,
in a simpler system of a single microsphere with gain, it has been
shown \cite{Lagendijk06}, that the scattering 
coefficients calculated within linear response lose 
their causality just at the point where the sphere crosses 
its lasing threshold. Applied to our random laser system, this means 
that the threshold for lasing within a spot of size $R_s$ is 
reached when the transport coefficient $-\gamma$, determined 
by the pump rate via the microscopic transport theory 
Refs. \cite{Frank06,Lubatsch09}, reaches the value given by 
Eq. (\ref{R_s})

\begin{figure}[t]
\begin{center} \scalebox{0.25}[0.25]{\includegraphics[clip]{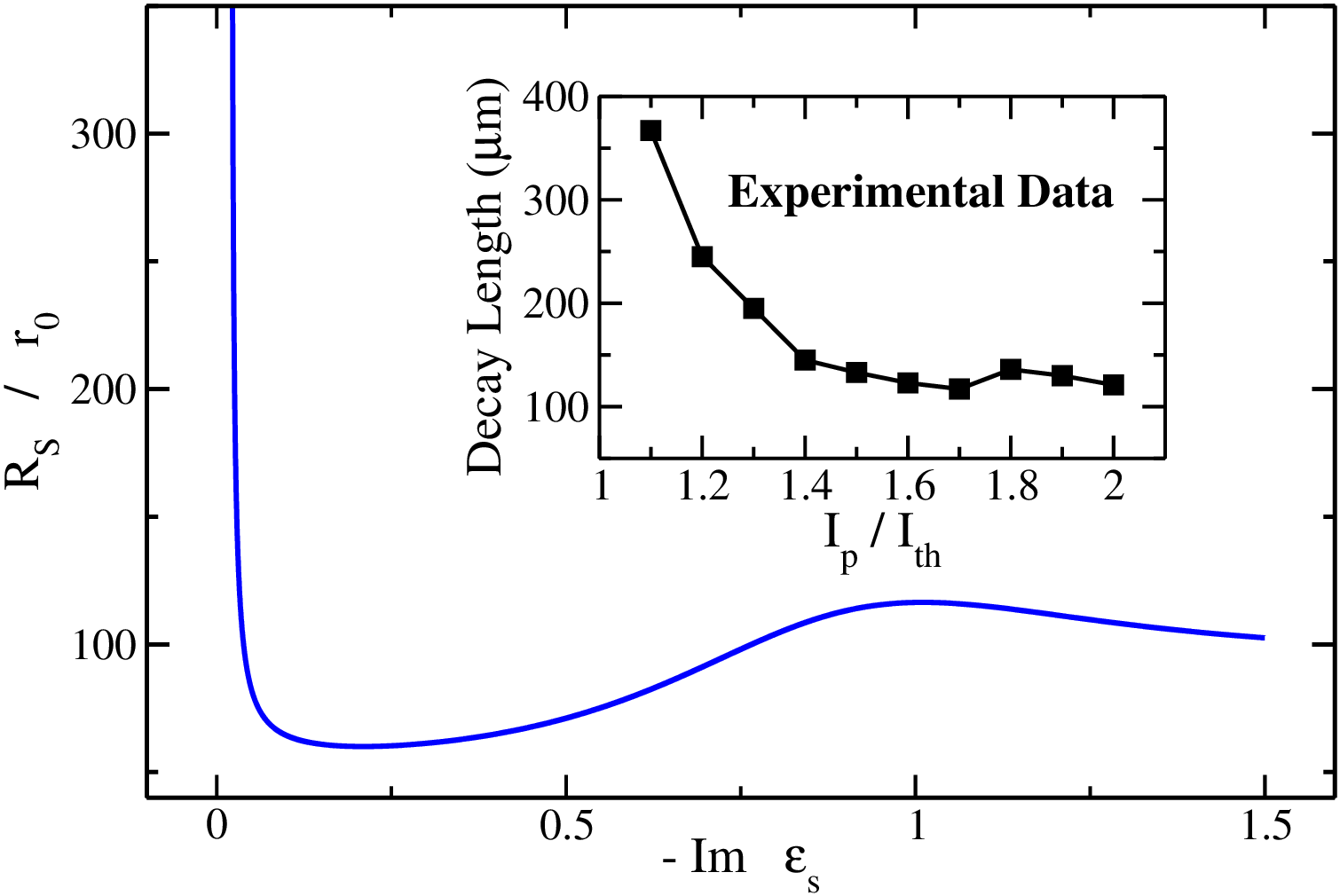}}\end{center}
\caption{
The spot size $R_{s}$, Eq. (\ref{R_s}) in units of the scatterer 
radius $r_0$, as 
obtained by causality considerations (see text) as a function
of the imaginary part of the dielectric constant of the 
scatterers.
The parameter values used are
$\epsilon_b =1$, $Re\epsilon_s=10$, scatterer filling fraction  $\nu=30\%$, 
light frequency $\omega / \omega_0 = 2.5$.
The ligth frequency is
$\omega_0 = 2\pi c/ r_0$ where $c$ is the vacuum speed of light.
The data in the inset are taken from 
Ref. \cite{Cao03} and refer to the spot size of the modes.
}
\label{Fig_0}
\end{figure}

In Fig. \ref{Fig_0} we show  
the numerical evaluation of the spot size  $R_{s}$
as a function of increasing  ${\rm Im} \epsilon_s$ for 
typical parameters, as given in the figure caption. 
The imaginary part of the dielectric constant
is a measure of external pumping, 
since the gain is given by the population inversion
of the laser. Therefore, 
larger  pumping yields higher inversion and leads to 
a larger  $ {\rm Im} \epsilon_s$.
The calculated spot size displays  a 
qualitative agreement with the experimental data \cite{Cao03} 
(spot size vs. pump intensity / threshold intensity) 
shown in the inset.

\section{Tranport Theory of Random Lasing}
\label{Sec_RL}

As remarked in section II, a stationary lasing state in a 
homogenaously pumped system is possible
only if the system is finite, so that surface loss effects can
compensate the gain in the medium. To avoid the causality problem, 
we consider here a three-dimensional 
random laser model with a homogeneously pumped,
active medium which extends infinitely in the (x,y) plane, but has a finite,
constant thickness $d$ in the z direction.
The laser-active material 
is  described  by the semi-classical laser rate equations, 
and the light intensity transport by a diffusion equation.
In particular, the rate equations for a four-level laser are
\begin{eqnarray}
\frac{\partial N_3}{\partial t} 
&=& 
\frac{N_0}{\tau_{P}}  - \frac{ N_3}{\tau_{32}} \\
\frac{\partial N_2}{\partial t} 
&=&  
\frac{N_3}{\tau_{32}}  -   \left(\frac{1}{\tau_{21}}
+ \frac{1}{\tau_{nr}}\right)N_2 -  
\frac{\left( N_2 -N_1\right)}{\tau_{21}} n_{ph} \\
\frac{\partial N_1}{\partial t} 
&=&    
\left(\frac{1}{\tau_{21}}+ \frac{1}{\tau_{nr}}\right)N_2  
+ \frac{\left( N_2 -N_1\right)}{\tau_{21}} n_{ph} 
-   \frac{ N_1}{\tau_{10}} \\
\frac{\partial N_0}{\partial t} 
&=&   
\frac{N_1}{\tau_{10}} - \frac{N_0}{\tau_{P}} \\
N_{tot} &=& N_0 + N_1 + N_2 + N_3,
\end{eqnarray} 
where $N_i=N_i(\vec r,t)$, $i=0,\ 1,\ 2,\ 3$ are the population 
number densities of the corresponding 
electron level ($i\in \{1\ldots 4\}$),
$N_{tot}$ is the total number of electrons participating in the 
lasing process,
$\gamma_{ij} \equiv 1 / \tau_{ij}$ are the transition 
rates from level $i$ to $j$, and $\gamma_{nr}$ is the non-radiative
decay rate of the laser level 2.
$\gamma_P \equiv 1 / \tau_P  $ is the transition 
rate due to homogeneous, constant, external pumping. Furthermore 
$n_{ph} \equiv N_{ph} / N_{tot}$ is the 
photon number density, normalized to $N_{tot}$. 
In the stationary limit (i.e. $\partial_t N_i = 0 $),  
the above system of equations can be solved for  
the population inversion $n_2=N_2/N_{tot}$ to yield
($\gamma_{32}$ and  $\gamma_{10}$ assumed to be large compared to 
all other rates) 
\begin{eqnarray}
\label{define_n_2}
n_2 =\frac{\gamma_P}{\gamma_P 
+ \gamma_{nr} + 
\gamma_{21}\left(n_{ph} +1\right)},
\label{n_2}
\end{eqnarray} 
The photon number density (light intensity), normalized to 
$N_{tot}$, $n_{ph}=N_{ph}/N_{tot}$, obeys 
the diffusion equation \cite{John04},
\begin{eqnarray}
\partial_t  n_{ph} = D_0 \nabla^2 n_{ph} 
+ \gamma_{21} (n_{ph} +1) n_2,
\end{eqnarray} 
where the last term on the r.h.s. describes the intensity 
increase due to stimulated and spontaneous emission, as 
described by the semi-classical laser rate equations.
Since in the slab geometry 
ensemble-averaged quantities are translationally invariant 
in the ($x,y$) plane, but not along the $z$ direction, a 
Fourier representation in the ($x,y$) plane in terms of 
$n_{ph}(\vec Q_{||},z)$, $n_2(\vec Q_{||},z)$ is convenient,
\begin{eqnarray}
\label{diff}
\partial_t  n_{ph} &=& - D_0 Q^2_{||} n_{ph} + D_0 \partial_z^2 n_{ph}\\
&+& \gamma_{21} \int \frac{d^2Q_{||}'}{(2\pi)^2} 
n_{ph}(\vec Q_{||}-\vec Q_{||}',z) n_2(\vec Q_{||}',z) 
+ \gamma_{21} n_2\nonumber
\end{eqnarray}
We now seek the photon density response
function $P(\vec Q_{||},z,\Omega)$, which describes the 
response of the photon density, $n_ph$, to the distribution of the 
population inversion, $n_2$, in order to determine the
transport coefficients. 
In the stationary case ($\partial_t n_{ph}=0$) 
and in the long-wavelength limit along the 
($x,y$) plane ($Q_{||}\to 0$), the $z$ derivative in 
Eq. (\ref{diff}) can be expressed without derivatives in terms of
$n_{ph}$ and $n_2$ only.
Plugging this back into Eq. (\ref{diff}) yields,
\begin{eqnarray}
\label{diff2}
\left[\partial_t  + D_0 Q^2_{||} + \frac{\gamma_{21}n_2}{n_{ph}}\right] 
n_{ph}(\vec Q_{\||,z},t) = \gamma_{21} n_2(\vec Q_{\||,z},t)
\end{eqnarray}
and, hence, after Fourier transform w.r.t. time, the diffusion
form of the density response function, 
\begin{eqnarray}
P_E(\vec{Q}_{||},z,\Omega)=
\frac{i\gamma_{21}}{\Omega
 +
i Q^2_{||} D_0 + i\xi^{-2}D_0
} \ ,
\label{Kernel_P_model}
\end{eqnarray}
where from Eq. (\ref{diff2}) the correlation length $\xi$ is defined as
the real, positive quantity,
\begin{eqnarray}
\label{xi}
\xi
=
\sqrt{
\frac
{D_0}
{
\gamma_{21}}\,\,\,\frac{n_{ph}}{n_2}
}.
\end{eqnarray}
As seen from Eq.  (\ref{Kernel_P_model}) 
the pole structure of $P_E$ in this finite-size, 
diffusive system is perfectly causal. 
The square of the correlation length  $\xi$
remains positive, indicating an effective 
loss out of a given $Q_{||}$ mode.
This is due to the loss of intensity 
at the surfaces.  Additionally, 
the mass term  becomes less and less significant as 
the  laser intensity in the sample builds up, 
because the relative population 
inversion clearly obeys $n_2 \le 1$ whereas the 
relative photon number is not restricted.

Since for homogeneous pumping 
the averaged photon density does not depend on 
$x$ or $y$, Eq. (\ref{diff}) simplifies in the stationary limit to
\begin{eqnarray}
\label{z-diffusion}
D_0 \partial^2_z n_{ph}   = - \gamma_{21}    (n_{ph} +1) n_2 
\end{eqnarray}
and $n_{ph} (z)$ is finally determined via Eq. (\ref{n_2}) by
the regular differential equation,
\begin{eqnarray}
\label{n_ph_of_z}
\partial^2_z n_{ph}  (z)  = 
-\frac{\gamma_{21} }{D_0}
\,\,\,
\frac{ (\gamma_P / \gamma_{21})}
{
1 + \frac{ (\gamma_P / \gamma_{21} ) }{n_{ph}(z)+1}
} \ .
\end{eqnarray}
Eqs. (\ref{n_ph_of_z}) (\ref{n_2}) and (\ref{xi}) 
comprise the complete description of the spatial 
photon density profile perpendicular to the lasing film 
and the intensity
correlation length (spot size) parallel to the film.

\begin{figure}[t]
\begin{center} \scalebox{0.5}[0.5]{\includegraphics[clip]{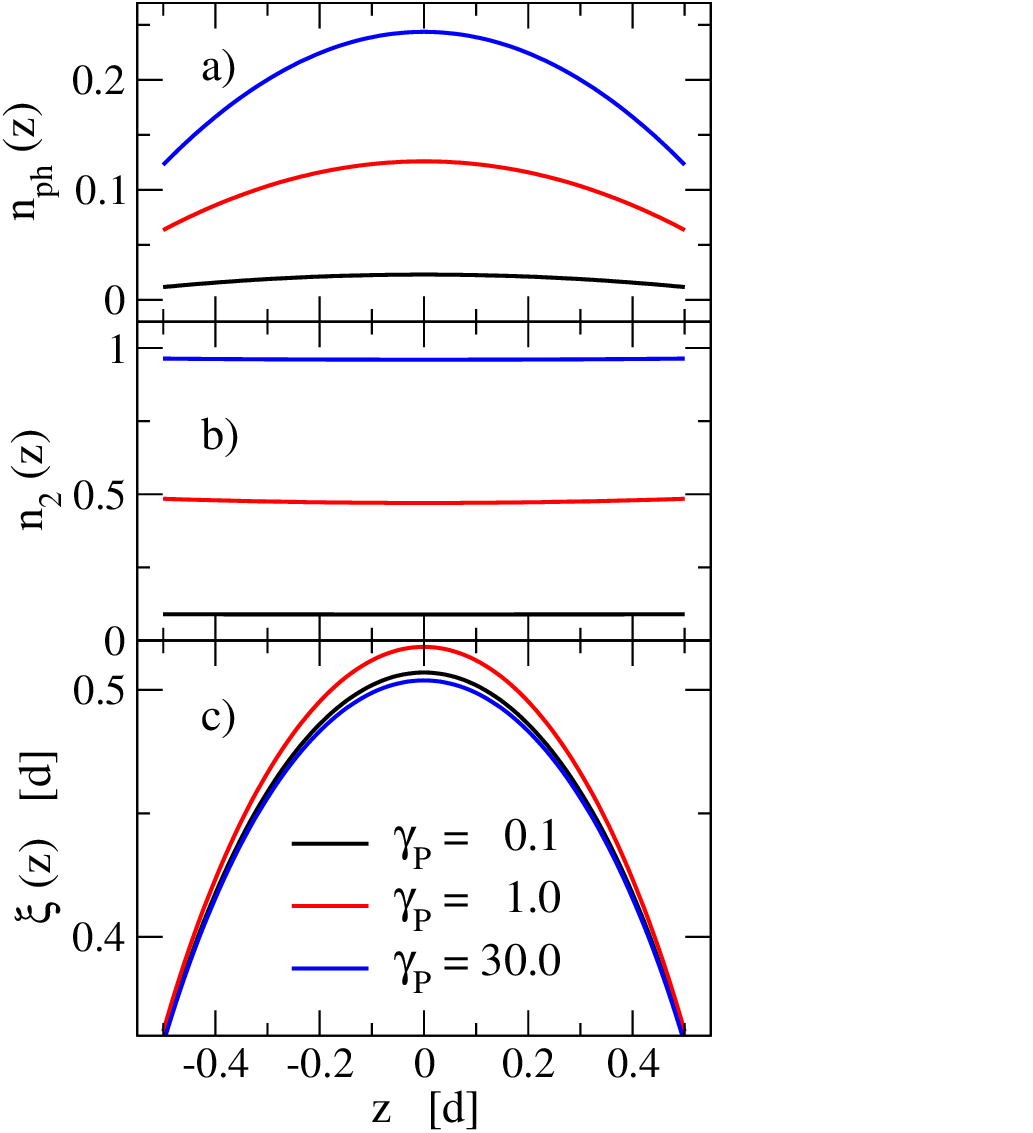}}\end{center}
\caption{
The following quantities are shown as a function of $z$ 
for different values 
of the pump rate  $\gamma_P$:
a) the photon number, which increases monotonically with increasing 
pump rate  and has its maximum in the film  center ($z=0$);
b) the population inversion, 
which is  inverse proportional to $n_{ph}$, see Eq. (\ref{define_n_2});
c) the correlation length (spot size), 
which clearly behaves non-monotonically with increasing pumping. 
For all panels the diffusion constant is $D_0 = 1 d^2\gamma_{21}$.
}
\label{Fig_1}
\end{figure}

\begin{figure}[t]
\begin{center} \scalebox{0.5}[0.5]{\includegraphics[clip]{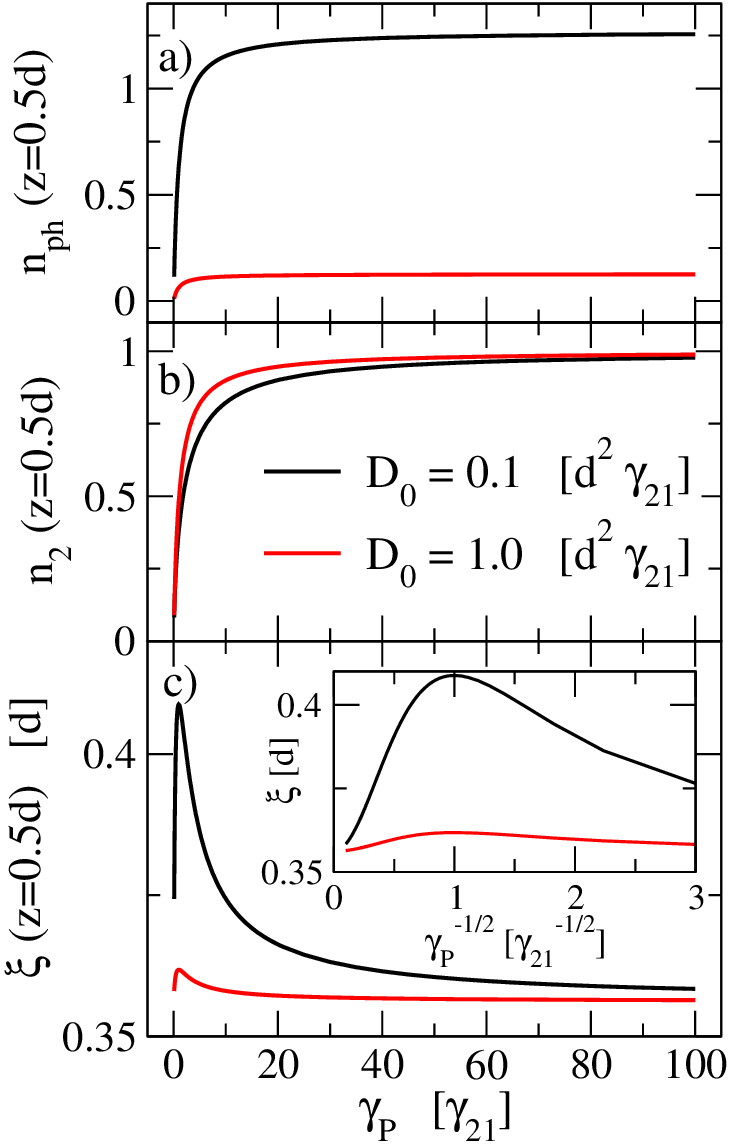}}\end{center}
\caption{
The figure shows the following quantities as 
a function of the pumping rate $P$ at the film surface ($z=\pm 0.5d$):
a) the photon number which displaying saturation for strong pumping;
b) the population inversion, also saturating;
c) the correlation length (spot size) showing
non-monotonic behavior. The inset of 
panel c) displays the $1/\sqrt{\gamma_P}$ dependence of the 
intensity correlation length $\xi$ 
on the pump rate above threshold ($\gamma_P>1$)
(see discussion in the text).
}
\label{Fig_2}
\end{figure}

Numerical evaluations of Eqs. (\ref{n_ph_of_z}), 
(\ref{xi}) and (\ref{n_2})  are 
shown in Figs. \ref{Fig_1} and \ref{Fig_2}. In Fig.  
\ref{Fig_1} the photon number $n_{ph}(z)$,  population 
inversion  $n_2(z)$ and  correlation length 
 $\xi (z)$ are shown as a function
of $z$  for different values of external pumping, 
characterized by the pumping rate $\gamma_P$. 
The value of the diffusion  constant was chosen to 
be $D_0= 1 d^2\gamma_{21}$, where 
$d$ is the width of the film. In panel a) of Fig. 
\ref{Fig_1} the photon number displays a monotonically 
increasing behavior with increasing pumping. The 
maximum of the intensity resides in the center of film 
($z=0$), since this is the position farthest from 
the boundaries, and therefore with lowest loss of 
intensity. The population inversion, Eq. 
(\ref{define_n_2}), behaves  inverse
to $n_{ph}(z)$, see Eq. (\ref{define_n_2}). 
In contrast to this rather expected behavior, 
the correlation length $\xi (z)$ as  given by Eq. (\ref{xi}) 
exhibits a non-monotonic behavior 
with increasing pumping. For pumping rates $\gamma_P < \gamma_{21}$ 
the correlation length increases but for 
pumping rates $\gamma_P > \gamma_{21}$, $\xi$ is decreasing. 
The equality between  $\gamma_P$ and $ \gamma_{21}$ marks 
the situation where electrons  are as fast excited  into 
the upper laser level as they relaxate to lower levels. 
Therefore this characterizes the lasing threshold. 
Available experimental data  \cite{Cao00,Cao03}
also report a decreasing behavior of the spot size above threshold.
Measurements of the intensity correlation length below threshold
have not yet been reported.

The same quantities  are shown Fig. \ref{Fig_2} as a function 
of external pumping $\gamma_P$ at the surface of the random laser.
Photon number and  population inversion  
both display saturation behavior. 
Panel c) of  Fig. \ref{Fig_2}, however,
exhibits the non-monotonic behavior of the correlation length.
This plot is to be directly compared to experimental data \cite{Cao03}, 
as e.g. shown in the inset of  Fig. \ref{Fig_1}.
There is a good qualitative and even quantitative agreement between 
calculated and measured spot size. The inset of panel c)
exhibits shows that the dependence of the spot size $\xi$ on the 
pump rate $\gamma_P$ above threshold ($\gamma_P$) is predicted to be 
\begin{equation}
\xi(\gamma_P)= \xi (\infty) + \alpha /\sqrt{\gamma_P} 
\end{equation}
with a proportionality constant $\alpha$. This result is open for 
further experimental tests.

\section{Conclusion}

We have discussed how a linear response theory for light transport,
including self-interference effects, in disordered media with linear
gain predicts threshold behavior of the intensity. Even 
more interestingly it also predicts a characteritic, average
radius of lasing modes, dictated by causality. 
We identify this length scale with the spot size of the random laser 
as measured in experiments \cite{Cao00,Cao03} and find 
qualitatively good agreement.
Further, we have proposed an analytical transport theory for random lasing
in finite systems. A finite system size is necessary for surface loss
to compensate the gain in the medium and, hence, to stabilize a
stationary lasing state. In particular, we consider a slab
geometry where in the medium the light intensity propagates diffusively and
the loss through the surfaces is included by appropriate boundary conditions.
The theory allows for the first time for an analytical calculation of 
the intensity correlation length of this system, describing the spatial 
extent of a mode (spot size).
The spot size is predicted to behave non-monotonously as a function of
external pumping,  i.e. increasing below and decreasing above the laser
threshold. A comparison with experiments reveals qualitativly good agreement.
Our prediction of its functional dependence on the pump rate is open to
further experimental tests.

Our future work will include solving the semi-analytical light
transport theory with self-interference contributions when 
the system is self-consistently coupled to the laser rate equations.

Useful discussions with H. Cao, P. Henseler, B. Shapiro, and C. M. Soukoulis
are gratefully acknowledged.
This work was supported in part by the Deutsche Forschungsgemeinschaft 
through grant no. KR1726/3 (R.F., J.K.) and FG 557.


\end{document}